\documentstyle[12pt]{article}
\begin{document}
\title{The 1/N Expansion and Long Range Antiferromagnetic Order}
\author{Maxim Raykin \thanks{Email: raykin@buphy.bu.edu}\\
{\normalsize {\it Department of Physics,
Boston University,
Boston, MA 02215, USA,}}\\
and \\
Assa Auerbach \thanks{Email: assa@phassa.technion.ac.il} \\
{\normalsize {\it Physics Department,
Technion-IIT,
Haifa 32000,
Israel.$^{{(a)}}$}}}
\maketitle
\begin{abstract}
The staggered magnetization of the Heisenberg antiferromagnet  in
two dimensions can be systematically approximated
by a 1/N expansion. Cancellation between self energy diagrams
leads to a Luttinger-like theorem for the ground state. We prove
(for a smooth enough self energy) that the long range order of mean
field theory ($N$=$\infty$) survives corrections to all orders of
$1/N$.  Divergences of this series provides a new route to the
disordered phases of quantum antiferromagnets.
\end{abstract}
PACS number(s): 75.10.Jm, 75.50.Ee, 67.40.Db
\newpage
In the study of quantum phase transitions, the
order-to-disorder transitions of the Heisenberg antiferromagnet at
zero temperature are particularly interesting. There is also hope
that  understanding such transitions may provide insight into the
electronic correlations of lanthanum cuprates where under low
doping, antiferromagnetism is replaced by superconductivity.

The ground state of the Heisenberg antiferromagnet in two dimensions
can  have either long range order or be disordered  by quantum
fluctuations \cite{sudip}. Disordering  can be induced by
frustrating longer range interactions, or perhaps by slight hole
doping as suggested by the phase diagram of lanthanum cuprates.
In either case, there are
theoretical difficulties in describing the transition itself. While
semiclassical spin wave theory works well deep in the ordered
phase, it assumes spontaneously broken symmetry, and fails when
the staggered magnetization becomes small. The continuum
approximation of the (2+1) dimensional non linear sigma model
relies on perfect short range antiferromagnetic correlations.
Near the transition however, the short range correlations
deteriorate considerably. This complicate matters, since one
needs to include field discontinuities (e.g. hedgehogs), and
consider interference effects between their Berry phases
\cite{haldane}.

The Schwinger Bosons (SB) large-N expansion \cite{aa} is a
rotationally symmetric formulation,
which in principle can treat both sides of the
transition \cite{RS}. The mean field theory (MFT)
describes the excitations as a free Bose gas of $N$ decoupled
flavors. Bose condensation in this system
is equivalent to long range spin order \cite{h}.
However, MFT is strictly valid only  at $N = \infty$, while the
physically interesting system is at $N$=2. A connection between
the two limits requires an understanding
of the $1/N$ series. Higher order corrections involve
interactions between SB which enforce the local
constraints. However, finite $N$ corrections  to the staggered
magnetization have not yet been evaluated. Before this could be
done, it is necessary to place the 1/N expansion on firmer
footing, i.e. to show that the higher order terms yield finite
and sensible results, which do not immediately destroy the mean
field ground state.  Currently, we do not know  whether the long
range order found in the MFT survives for  {\em any} $N<\infty$.

This paper specifically addresses this concern. We  prove  a
theorem which establishes the 1/N expansion as a consistent
approach for the ground state of finite $N$ systems, starting
from the MFT. Under a condition that the self energy is sufficiently
smooth at the ordering wavevectors,  we prove that {\em if there is
long range order in the MFT, the spontaneous staggered magnetization
does not vanish to all orders of the 1/N expansion.}
The proof uses
a cancellation between self energy diagrams and their tadpole
counterparts, a feature special to the 1/N expansion.

The result is reminiscent of (but not equivalent to) Hugenholtz and
Pines' self energy condition for Bose condensed liquids \cite{HP}.
It is closer in spirit to Luttinger's theorem for
Fermi liquids \cite{l}. The spontaneous staggered magnetization is
analogous to Luttinger's Fermi surface. (Both appear as a
discontinuity in the occupation number.) In Luttinger's theorem,
under a similar condition on the self energy, the Fermi surface
discontinuity survives at each order in perturbation theory.
Pushing this analogy further, we shall propose that the vanishing
of the staggered magnetization at finite $N$ may formally resemble
one of the known Fermi surface instabilities.

For simplicity we discuss the nearest neighbor SU(N) Heisenberg
antiferromagnet. The proof actually utilizes only general
features of this model, and thus it is readily extendable to more
general hamiltonians. The spins are represented by $N$ SB per site
$a^\dagger_{i,-m_{max}}\ldots a^\dagger_{i,m_{max}}$
where $m_{max}=(N-1)/2$, and the Hamiltonian is given by
\begin{equation}
{\cal H}~=-{J\over N}\sum_{{\langle i,j \rangle} } (a^\dagger_{im}
a^\dagger_{j m})
(a_{i m'} a_{jm'})~- h \sum_{im}~ma^\dagger_{im} a_{im},
\label{1}
\end{equation}
${\langle i,j \rangle}$ are nearest neighbor bonds on the square
latice. Summation over repeated indices is implicit, unless
specified otherwise. The Hilbert space is constrained
by the fixed SB number
$a^\dagger_{im}a_{im}~=Ns$, at each site. $h$ is an infinitesimal
ordering field. The generators of SU(N) are given by
$a^\dagger_{im}a_{im'}$, where we use conjugated representations
on opposite sublattices. For N=2,
Eq. (\ref{1}) is equivalent (up to a constant) to the   Heisenberg
model ${\cal H}=J\sum_{{\langle i,j \rangle}}{\bf S}_i\cdot
{\bf S}_j-h\sum_i (-1)^{\bf i}S^z_i$ \cite{aa}.

Following the standard procedure \cite{aa}, the partition function
can be written as a coherent states path integral. Hence one
introduces real local fields $\lambda_i$ to impose the constraints,
and  Hubbard Stratonovich fields $Q_{ij}$ to decouple the quartic
interactions.

After integrating out the SB field, we are left with
\begin{equation}
Z(h)~= \int {\cal D} (\lambda Q) ~\exp\left[ - N{\cal S}
(\lambda,Q,h)\right]
\label{2}
\end{equation}
The explicit expression for the action ${\cal S}$ can be found in Ref.
\cite{aa}. Following \cite{h}, we consider the case of zero
temperature and large, but finite volume. The staggered
magnetization is given by
\begin{equation}
M ~= \lim_{h\to 0^+}~\lim_{\beta\to \infty}
{1\over 2\beta V}  {\rm Tr}~ m G(k,m)
\label{3}
\end{equation}
where $G$ is the full Green function of SB. It is
given in the Nambu notation as a $2\times 2$ matrix with normal
and anomalous components $G_{nn'}(k,m)=\langle A_{nkm}
A^{\dag}_{n'km} \rangle$, where $A_{km}=(a_{km},a_{-km}^\dagger)$.
$\beta$ is the inverse temperature and $V$ is the volume (number of
lattice sites). We denote $k=({\bf k},\omega)$, where ${\bf k}$ and
$\omega$ are lattice momentum and Matsubara frequency respectively.
${\rm Tr}$ includes a trace over $k,m$ and the Nambu indices. $G$
is evaluated by summing all one-particle diagrams generated by the
large-N expansion of Eq. (\ref{2}).

Let us briefly review the mean field results which were derived
previously \cite{aa,h}. At large $N$,  (\ref{2}) is
dominated by the saddle point ${\bar \lambda},~{\bar Q}$, and $G$
is approximated by the mean field Green function $G_0$,
\begin{eqnarray}
G_{0}(k,m)~=\pmatrix{ {\bar \lambda} - i \omega - hm & 4 {\bar Q}
\gamma_{{\bf k}} \cr
4 {\bar Q} \gamma_{{\bf k}} & {\bar \lambda} + i\omega - hm }^{-1}
\label{4}
\end{eqnarray}
where $\gamma_{\bf k}~={\frac{1}{2}}(\cos k_x + \cos k_y )$. The
poles of $G_0$ are at the SB frequencies
\begin{equation}
\omega_{{\bf k},m} =c\sqrt{\Delta _h^2 + {{\bar \lambda} \over
4 {\bar Q}^2} h( m_{max} - m ) + 2 ( 1-\gamma_{\bf k}^2 ) }
\label{5}
\end{equation}
where $c = \sqrt{8} {\bar Q}$ and $\Delta_h = c^{-1} \sqrt{
{\bar \lambda}^2 - 16 {\bar Q}^2 - 2 h {\bar \lambda} m_{max}}$.
$\omega_{\bf k}$ is  minimized at the two points
${\bf k}_c=(\vec 0,\vec\pi)$. At those momenta, for  $m=m_{max}$,
the excitation gap is $c \Delta_h$.
Solving the mean field equations
yields $\Delta_h ~= \sqrt{2}/(NV(s-0.1966\ldots))$. In the
thermodynamic limit, the SB with $m=m_{max}$ and
${\bf k}={\bf k}_c$ contribute macroscopically
to the momentum sum, i.e.
they undergo {\em Bose condensation}. This condensate is the only
term, which survives the cancellation between positive and negative
$m$'s in (\ref{3}), yielding the mean field staggered magnetization
\begin{equation}
M_0~={N - 1 \over 2} {\sqrt{2} \over V \Delta_h}~={N (N-1)
\over 2} (s-0.1966\ldots)
\label{6}
\end{equation}
which for $N=2$ agrees with  spin wave theory \cite{h}.

The higher order $1/N$ corrections to $G$ are described by diagrams
which include lines for $G_0$ interacting via propagators $D$
(defined later) which are depicted as wiggly lines. A diagram which
involves $L$ loops (traces of products of $G_0$)  and  $P$
propagators, is of order $(1/N)^p$ where $p=P-L$. One must exclude
all diagrams which include the segments shown in Fig. 1.
As shown in \cite{aa,ra}, this leads to the fulfillment of the SB
constraints at each order of $1/N$ separately.

As in MFT, a nonzero staggered magnetization is related to the
divergence of the number of SB with $m=m_{max}$ at
${\bf k}={\bf k}_c$.
On the other hand, strictly at $h=0$, MFT is SU(N) rotationally
invariant, so the Bose condensation is equally shared among the
different $m$ flavors and the gap becomes $\Delta_0 =
N\Delta_{h\ne 0}$. Henceforth we shall set $h=0$, and have exact
degeneracy between the different flavors $\omega_{{\bf k},m}$.
Thus, long range order is associated with Bose condensation of
all flavors at ${\bf k}_c$.

The self energy $\Sigma(k)$ (also a $2\times 2$ matrix)
is related to $G$ by the Dyson  equation
$G^{-1}= G_0^{-1}-\Sigma $.
In order to proceed we must make an important assumption on
the smoothness of the self energy  near the condensate momenta:
\begin{equation}
\lim_{k\to k_c} \left|\Sigma (k)-\Sigma (k_c)\right| =
O\left(|k-k_c|^{2-\delta}\right),
{}~~~~\delta< 1
\label{16}
\end{equation}
where $k_c\equiv ({\bf k}_c,0)$.
$\Sigma$ should exhibit rotational symmetry about $k_c$ as a
consequence of the asymptotic ``Lorentz  invariance''
of $G_0$ near $k_c$. (The SB dispersion vanishes linearly at
${\bf k}_c$.) We have verified that the leading order
self energy is smooth at $k_c$ (i.e. obeys (\ref{16}) with
$\delta=0$). We argue that  the smoothness assumption is plausible
for models which have no pathology in the density of low
excitations. For such models, the integrations in  $\Sigma$ are
uniformly convergent for all external momenta.

However,  we have not {\em proven} Eq. (\ref{16})
to all orders in $1/N$, and we must regard it as an
{\em assumption}; one  which requires a separate justification
for any particular model.

The number of SB with momentum ${\bf k}_c$ is $n_{{\bf k}_c}=
(2\beta)^{-1}\sum_{\omega} {\rm Tr} G({\bf k}_c,\omega)$, where Tr
traces over Nambu indices. This number diverges as $n_{{\bf k}_c}
\sim V$ if $\det[G^{-1}(k_c)]\sim V^{-2}$. We use the Dyson
equation and MFT relation $\Delta_0 \sim V^{-1}$ to state that
\begin{equation}
\Delta' \equiv  \Sigma_{11}(k_c) - \gamma_{{\bf k}_c}
\Sigma_{12} (k_c) = O( V^{-2})~\leadsto ~M\ne 0
\label{8}
\end{equation}
i.e. if the quantity $\Delta'$ vanishes rapidly enough in the
thermodynamic limit, the ground state has long range order. It
may be shown, that $\Delta'(\vec 0)=\Delta'(\vec\pi)$.

{\em Theorem:} Under condition (\ref{16}), Eq. (\ref{8}) holds to
all orders in the $1/N$ series.

{\em Proof:}
The self energy is decomposed into two parts
\begin{equation}
\Sigma~={\tilde\Sigma} + \Sigma^{tad}
\label{9}
\end{equation}
where $\Sigma^{tad}$ is the {\em single} tadpole diagrams
(see Fig. 2), and ${\tilde\Sigma}$ are all the remaining diagrams.
Although $\tilde\Sigma$ and $\Sigma^{tad}$ are expected to be of
$O(1)$ separately, we shall show that at $k=k_c$ the $O(1)$
contributions precisely cancel  in (\ref{8}) leaving us with terms
of $O(V^{-2})$. {\em Note that in contrast to perturbation theory,
the first and second terms in Fig. 2. have
a different number of vertices, but are of the same order in
$1/N$}. This enables the cancellation mechanism function at
each order separately.

The rest of this discussion contains unavoidable technical
details. The set of auxiliary fields is denoted by $(\lambda_j,
\Re Q_{s,j},\Re Q_{d,j},\Im Q_{s,j},\Im Q_{d,j})$.
$\lambda_j$ couples
to the local Boson density  and $\Re (\Im) Q_{s(d),j}$ couple to
the bilinear forms  $\sum_{e=e_x,e_y} \eta_{e}^{s(d)} [a_j^\dagger
a_{j+e}^\dagger +(-) a_j a_{j+e}]$, where $\eta^s_{e_x}=\eta^s_{e_y}
=\eta^d_{e_x}=1$ and $\eta^d_{e_y}=-1$. We define $2\times 2$
vertices $\hat v^\alpha$ which connect
between a field $\alpha$ and two
$G_0$'s. Thus a zero momentum field $\alpha$ is coupled to the form
$\sum_{\bf k} A^{\dag}_k \hat v^\alpha_{{\bf k}} A_k$, where
$\hat v_{{\bf k}}^1= iI/2$,
$\hat v_{{\bf k}}^{2,3}=\sigma^x(\cos k_x \pm
\cos k_y)$, and $\hat v_{{\bf k}}^{4,5}=
i\sigma^y(\cos k_x \pm \cos k_y)$.
Using $\hat v^\alpha$,  we can explicitly write
$\Sigma^{tad}(k_c)$ (see Fig. 2) as
\begin{eqnarray}
\Sigma^{tad}(k_c)&=& 2 N \hat v^\alpha_{{\bf k}_c}
D^{\alpha,\alpha'}(0) \sum_k {\rm Tr} \left[
\hat v_{{\bf k}}^{\alpha'} G_0(k) R(k) G_0(k) \right] \nonumber\\
R~&=& {\tilde\Sigma} +\Sigma G \Sigma
\label{10}
\end{eqnarray}
where $\sum_k = {1\over V}\sum_{\bf k} \int {d \omega \over 2 \pi}$.
It may be seen, that only $\alpha, \alpha' = 1,2$ give nonvanishing
contribution to this formula.

$\Sigma(k)$ may be expanded as $\Sigma(k) = \Sigma_{\alpha}(k)
\hat u^{\alpha}_{{\bf k}}$, where summation over $\alpha$ runs from
1 to 3, $\hat u_{{\bf k}}^{1,2} = \hat v_{{\bf k}}^{1,2}$ and
$\hat u^3_{{\bf k}}=
\sigma^z$. The coefficients of expansion satisfy the relation
$\Sigma_{\alpha}(\vec 0,0) = \Sigma_{\alpha}(\vec\pi,0)$ and
$\Sigma_3(k_c)=0$. The same expansion with $\hat u^\alpha$
holds for $R(k)$. $\Delta'$ can now be written as $\Delta'=
f^{\alpha} \Sigma_{\alpha} (k_c)$, where $f^1=i/2$, $f_2=-2$ and
$f^{\alpha}=0,~\alpha>2$.

The propogator in the 1/N expansion is given by the matrix
$D~={1\over N} \left(\Pi_0 -\Pi\right)^{-1}$, where
\begin{eqnarray}
\Pi_0^{\alpha,\alpha'} (q)~&=& \delta_{\alpha,\alpha'} ( 1 -
\delta_{\alpha,1} ) {4 \over J}\nonumber\\
\Pi^{\alpha,\alpha'}(q)~&=& 2 \sum_k
{\rm Tr}\left[ \hat v_{{\bf k},{\bf k+q}}^{\alpha} G_0 (k+q)
\hat v_{{\bf k+q},{\bf k}}^{\alpha'} G_0(k) \right]
\label{11}
\end{eqnarray}
Using the relation $ND\Pi = -1 +ND\Pi_0 $, we find that at $k=k_c$,
${\tilde\Sigma}(k_c)$ is cancelled on the right hand side of Eq.
(\ref{9}) and we obtain
\begin{eqnarray}
\Sigma_\alpha(k_c)~&=& -\left[\Sigma(k_c)G(k_c)\Sigma(k_c)
\right]_\alpha + {4N\over J}D^{\alpha,2}(0)R_2 (k_c)\nonumber\\
&&~+ 2 N D^{\alpha,\alpha'}(0) \sum_k
{\rm Tr} \Biggl\{ \hat v_{{\bf k}}^{\alpha'}
G_0(k) \hat u_{{\bf k}}^{\alpha''} G_0(k)~\nonumber\\
&&~~~\times \left[ \left({\Sigma(k)\over 1-G_0(k)\Sigma(k)}
\right)_{\alpha''} - \left({\Sigma(k_c)\over 1-G_0(k_c)\Sigma(k_c)}
\right)_{\alpha''} \right] \Biggr\}
\label{12}
\end{eqnarray}
where we have used the fact that $\Sigma^{tad}_{\alpha''}$ is
independent of momentum.

The Bose condenstation of $G_0(k_c)$ gives rise to the divergence
of $\Pi(q=0)$. Extracting the volume divergences in (\ref{11})
yields $\Pi^{\alpha,\alpha'}(0) = f^{\alpha} f^{\alpha'} (aV^2
+ b V) + {\tilde P}^{\alpha,\alpha'}$ where
$a$, $b$, and ${\tilde P}^{\alpha,\alpha'}$ are
independent on volume. We see that both order $V^2$ and order $V$
factorize, reflecting the emergence of a disconnected part in the
correlation function due to Bose condensation. Denoting $P=\Pi_0-
{\tilde P}$ and inverting the polarization matrix we obtain
the propagator to order $V^{-2}$:
\begin{equation}
N D^{\alpha,\alpha'} (0) = (P^{-1})^{\alpha,\alpha'} -
{ (P^{-1})^{\alpha,\beta}f^{\beta}f^{\beta'} (P^{-1})^{\beta',
\alpha'} \over f^{\gamma} (P^{-1})^{\gamma,\gamma'}f^{\gamma'}}
-{1\over a V^2} { (P^{-1})^{\alpha,\beta}f^{\beta}f^{\beta'}
(P^{-1})^{\beta',\alpha'} \over
\left[f^{\gamma} (P^{-1})^{\gamma,\gamma'}f^{\gamma'}\right]^2}
\label{15}
\end{equation}
We now expand $\Sigma$ in a power series of $1/N$: $\Sigma~
=\sum_p N^{-p}\Sigma^{(p)}$. We shall prove Eq. (\ref{8})
by induction. Assume that Eq. (\ref{8}) holds for
$\Sigma^{(p)},p\le {\bar p}$. We take $\Sigma^{({\bar p}+1)}
_{\alpha}$ on the left hand side of
Eq. (\ref{12}), and multiply both sides by the vector $f^{\alpha}$.
Using the Dyson equation for $G$, one can show, that $f^\alpha
[\Sigma G \Sigma]_\alpha$ is proportional to $\Delta'$, which,
however, should be calculated with $\Sigma^{p\le {\bar p}}$.
Therefore, this term yields $O(V^{-2})$. Then, using
(\ref{15}), the terms of $O(1)$ in $D$ get cancelled by multiplying
them on the left by $f^\alpha$, leaving us with an overall factor
of $O(V^{-2})$. We must still show  that the second factor
in the third term of (\ref{12}) is not divergent. Since the summand
diverges as
$(k-k_c)^{-2}$, (2 powers of the phase space minus four powers from
$G_0$), the  momentum sum will converge if the self energy
obeys condition (\ref{16}). Thus, we have shown that Eq. (\ref{8})
holds to all orders in $1/N$. {\em Q.E.D.}

We note that it is crucial for the cancellation, described above,
that the constraint has a {\em local} character and enforced by a
fluctuating field. Indeed, this cancellation does not take place,
if the constraint is imposed only on average by a static
chemical potential. On the other hand, our proof can be readily
extended to different spin models with constrained Hilbert
spaces. In particular it applies to  the t'-J model, a
semiclassical approximation to holes in the quantum antiferromagnet
\cite{al}. Also, a simpler version of this theorem applies to the
long range order in Resonating Valence Bonds states \cite{lda},
using a large-N expansion  of the Gutzwiller projection \cite{ra}.

In practical terms, this theorem sets the foundation for
investigating the disordering transition using the $1/N$ expansion
of the self energy. We can propose two scenarios for the
disordering mechanism at finite $N$: (i) Coupling of spins to soft
charge fluctuations (holes) can give rise to  violation of
(\ref{16}), i.e. a breakdown of our theorem and
a destruction of long range order. This scenario is analogous to
the one dimensional Luttinger model where the
Fermi surface discontinuity vanishes due to the large density of
low excitations. (ii) The divergence of $V^2\sum_pN^{-p}
\Sigma^{(p)}\to \infty$ may be detected in a partial resummation
scheme. (Tadpole countertems must be properly included, as shown
above.) A divergence for example in nested diagrams, formally
resembles the Cooper channel (superconductivity) instability in
a Fermi liquid.

In summary, we have analyzed the corrections to the mean field
ground state staggered magnetization of the  two dimensional
antiferromagnetic Hei\-sen\-berg model. We found an important
cancellation mechanism  between self energy diagrams. This
establishes that the $1/N$ expansion for the order parameter is
a consistent asymptotic approach for finite $N$ models. It is
similar to perturbation theory about a non interacting Fermi
surface. We argue that the quantum disordering transition may
be detected as a breakdown of the
assumptions of this theorem, or a divergence in the $1/N$ series.
These possibilities are worth further investigations.

We thank  B.I. Halperin, L.P. Pitaevskii and C.M. Bender for
helpful comments.  This paper was supported in part
by grants from the US-Israel Binational Science Foundation, the
Fund for Promotion of Research at the Technion,
and by the US Department of Energy grant No. DE-FG02-91ER45441.

\vfill\eject

\vfill\eject

\section*{Figure Captions}

\begin{description}

\item[Fig. 1:] Forbidden segments in 1/N expansion diagrams.
Solid lines represent mean field Green
functions $G_0$ and wavy lines -- propagators of auxiliary
fields $D$ (see text).

\item[Fig. 2:] Diagrammatic representation of Eqs. (\ref{9},
\ref{10}) for the self energy.  ${\tilde\Sigma}$ are all diagrams
except the single tadpole diagrams. The cancellation between
${\tilde\Sigma}$ and the tadpole diagrams allows the staggered
magnetization to survive finite $N$ corrections.

\end{description}

\vfill\eject


\begin{thebibliography}{99}

\item[$^{(a)}$] Also at the Department of Physics, Boston
University, Boston, MA 02215.
\bibitem{sudip} E. Manousakis, Rev. Mod. Phys. {\bf 63}, 1 (1991),
and references therein.
\bibitem{haldane} F.D.M. Haldane,   Phys. Rev. Lett. {\bf 61}, 1029
(1988).
\bibitem{aa} D. P. Arovas and A. Auerbach, Phys. Rev. B {\bf 38},
316 (1988); A. Auerbach and D. P. Arovas, J. Appl. Phys. {\bf 67},
5734 (1990).
\bibitem{RS}  Topological Berry phases, however, may be
missing in the 1/N expansion of the disordered phase: N. Read and
S. Sachdev, Nucl. Phys. {\bf B316}, 609 (1989); Phys. Rev. Lett.
{\bf 62}, 1694 (1989); Phys. Rev. B {\bf 42}, 4568 (1990).
\bibitem{h} J. E. Hirsch and S. Tang, Phys. Rev. B {\bf 39}, 2850
(1989); M. Takahashi, {\it ibid.} {\bf 40}, 2494 (1989); S. Sarker,
C. Jayaprakash, H. R. Krishnamurthy and M. Ma, {\it ibid.} {\bf 40},
5028 (1989).
\bibitem{HP} N.~M.~Hugenholtz and D. Pines, Phys. Rev. {\bf 116},
489 (1959). HP use perturbation theory for an ordinary Bose liquid,
while here we use the special properties of the
$1/N$ expansion to treat a locally constrained
Schwinger bosons system.
\bibitem{l} J. M. Luttinger, Phys. Rev. {\bf 119}, 1153 (1960).
\bibitem{ra} M. Raykin and A. Auerbach, Phys. Rev. B {\bf 47}
(1993), and unpublished.
\bibitem{al} A. Auerbach and B.E. Larson,  Phys. Rev. Lett.
{\bf  66}, 2262 (1991). For previous discussions of the t'-J model
and superconductivity: P.B. Wiegmann,  Phys. Rev. Lett. {\bf  60},
821 (1988); P.A. Lee, {\em ibid.} {\bf  63}, 680 (1989).
\bibitem{lda} S. Liang, B. Doucot and P.W. Anderson,
Phys. Rev. Lett. {\bf  61}, 365 (1988).
\end{thebibliography}
\end{document}